\documentclass[english]{article}
\usepackage[T1]{fontenc}
\usepackage[latin9]{inputenc}
\usepackage{graphicx}
\usepackage{amssymb}

\makeatletter
\newcommand{\lyxaddress}[1]{
\par {\raggedright #1
\vspace{1.4em}
\noindent\par}
}

\makeatother

\usepackage{babel}

\begin{document}

\title{Generalization of correlated electron-ion dynamics from nonequilibrium
Green's functions}

\author{Yu Wang}

\maketitle

\lyxaddress{\emph{Atomistic Simulation Centre, School of Mathematics and Physics,
Queen's University Belfast, Belfast BT7 1NN, United Kingdom}}

\lyxaddress{E-mail: yu.wang@qub.ac.uk}

\begin{abstract}
We present a new formulation of the correlated electron-ion dynamics
(CEID) by using equations of motion for nonequilibrium Green's functions,
which generalizes CEID to a general nonequilibrium statistical ensemble
that allows for a variable total number of electrons. We make a rigorous
connection between CEID and diagrammatic perturbation theory, which
furthermore allows the key approximations in CEID to be quantified
in diagrammatic terms, and, in principle, improved. We compare analytically
the limiting behavior of CEID and the self-consistent Born approximation
(SCBA) for a general dynamical nonequilibrium state. This comparison
shows that CEID and SCBA coincide in the weak electron-phonon coupling
limit, while they differ in the large ionic mass limit where we can
readily quantify their difference. In particular, we illustrate the
relation between CEID and SCBA by perturbation theory at the fourth-order
in the coupling strength. \newpage{} 
\end{abstract}

\section{Introduction}

One of the most fundamental problems in molecular electronics is to
understand the inelastic scattering effects of atomic vibrations on
transmitted electrons. These effects have been extensively studied
both experimentally and theoretically in recent years \cite{Review}.
The interplay between electronic and nuclear dynamics in atomic-scale
devices influences not only the device characteristics, e.g. inelastic
current-voltage spectroscopy \cite{Horsfield-Bowler-Fisher-Todorov-Sanchez-2005},
but also the device stability due to local heating within the junction
\cite{Todorov-1998}.

Nonequilibrium Green's function (NEGF) theory \cite{Keldysh,Haug-Jauho-Book-1996,Rammer-Book-2007}
provides a systematic framework for describing the effects of the
coupling between transmitted electrons and atomic vibrations \cite{Caroli-Combescot-Nozieres-SaintJames-1972,Hyldgaard-Hershfield-Davies-Wilkins-1994,Galperin-Ratner-Nitzan-2004,Frederiksen-Paulsson-Brandbyge-Jauho-2007}.
For weak electron-phonon coupling, a well-known approximation for
evaluating the dressed Green's function is the self-consistent Born
approximation (SCBA) \cite{Hyldgaard-Hershfield-Davies-Wilkins-1994,Galperin-Ratner-Nitzan-2004,Frederiksen-Paulsson-Brandbyge-Jauho-2007}
which sums only noncrossing diagrams in the diagrammatic perturbation
expansion of the Green's function. Because the percentage of noncrossing
diagrams decreases quickly with increasing order in the coupling strength,
SCBA breaks down at strong electron-phonon coupling \cite{Goodvin-Berciu-Sawatzky-2006}.

Alternatively, correlated electron-ion dynamics (CEID) \cite{Horsfield-Bowler-Fisher-Todorov-Sanchez-2005,Horsfield-Bowler-Fisher-Todorov-Sanchez-2004,Horsfield-Bowler-Ness-Sanchez-Todorov-Fisher-Review-2006}
has been developed for describing the effects of the electron-ion
correlation and interaction on the inelastic dynamics of the electrons
and nuclei. CEID, as an extension of molecular dynamics, reinstates
the electron-ion correlation and the quantum nature of nuclei in order
to take account of energy exchange between electrons and nuclei reasonably.
So far CEID has been applied to a wide range of transport properties
of atomic wires, including inelastic current\emph{-}voltage spectroscopy
\cite{Horsfield-Bowler-Fisher-Todorov-Sanchez-2005}, the calculation
of local heating (and its signature on the current) in real time when
combined with electronic open boundaries \cite{McEniry-Bowler-Dundas-Horsfield-Sanchez-Todorov-2007},
and the non-conservative nature of current-induced forces \cite{Dundas-McEniry-Todorov-2009}.
Recently, a comparison between CEID and the NEGF method in SCBA has
been made both numerically and analytically for steady state transport
\cite{McEniry-Frederiksen-Todorov-Dundas-Horsfield-2008}. However,
there are two restrictions on the original CEID discussed above. First,
it is assumed that the electron-ion system is described in terms of
an ensemble with a fixed total number of electrons. Second, the original
CEID methodology lacks a systematic scheme to improve its accuracy.

In this paper, we develop a new formulation of CEID by using equations
of motion for a set of nonequilibrium Green's functions which are
closely linked to the dynamical variables in the CEID method. To illustrate
this idea, we consider a model system of noninteracting electrons
linearly coupled to a quantum oscillator. The motivation behind this
effort is to lift the restrictions on the original CEID. We attempt
to make a rigorous connection between CEID and diagrammatic perturbation
theory, so as to quantify the key CEID approximations in diagrammatic
terms, and, in principle, to be able to improve them. Moreover, in
the framework of NEGF, the scope of CEID can be readily extended to
a general nonequilibrium ensemble with a variable total number of
electrons. We compare analytically CEID with SCBA for a general nonequilibrium
state in the time domain, thus extending the previous comparison \cite{McEniry-Frederiksen-Todorov-Dundas-Horsfield-2008}
for a steady state in the energy domain.

The paper is organized as follows. In the next section, we present
the new formulation of CEID for the model system by using equations
of motion for a set of nonequilibrium Green's functions. In section
\ref{sec:Comparison}, CEID is analytically compared with SCBA for
the model system at the fourth-order in the coupling strength and
in two specific limits: weak electron-phonon coupling limit and large
ionic mass limit. To evaluate the Green's functions for the system
of electrons coupled to a classical oscillator, a classical version
of Wick's theorem is introduced in the appendix. Finally, conclusions
are drawn in section \ref{sec:Conclusions}.

\section{Model and formulation\label{sec:Model-and-formulation}}

We consider an infinite open system of noninteracting electrons linearly
coupled to a single quantum oscillator. The electrons are described
in terms of the second-quantized field operators $\Psi(r)$ and $\Psi^{+}(r)$.
The Hamiltonian of the system then takes the form

\begin{eqnarray}
H & = & \int dr\Psi^{+}(r)\left(-\frac{\hbar^{2}}{2m}\bigtriangledown^{2}+V(r)\right)\Psi(r)+\left[\frac{P^{2}}{2M}+\frac{1}{2}KX^{2}\right]\nonumber \\
 &  & -X\int drF(r)\Psi^{+}(r)\Psi(r)\label{eq:model-Hamiltonian}\end{eqnarray}
 with $X=R-R_{0}$. The first two terms constitute the free-particle
Hamiltonian $H_{0}$, and the last term describes the electron-phonon
interaction $H^{i}$. Here $V(r)$ is the lattice potential and $F(r)$
is the electron-phonon coupling strength. $R_{0}$ and $K$ are the
equilibrium position and the spring constant of the harmonic oscillator
respectively.

In the absence of electron-phonon interaction $H^{i}$ (taken to exist
at $t=-\infty$), the unperturbed electron subsystem was settled in
the Landauer steady state, which is characterized by two sets of one-electron
states, i.e. Lippmann-Schwinger scattering states, $\left\{ \left|\Phi_{i\alpha}\right\rangle \right\} $
$(\alpha=1,2)$ with occupancies $f_{i\alpha}$ set by the battery
terminals \cite{Todorov-Briggs-Sutton-1993,Todorov-2002}. In the
$\left|\Phi_{i\alpha}\right\rangle $-representation, the statistical
operator of the unperturbed system is thus taken to be

\[
\rho_{0}=\frac{1}{Z}\exp\left\{ -\beta\left[\left(\frac{P^{2}}{2M}+\frac{1}{2}KX^{2}\right)+\sum_{i}\sum_{\alpha=1}^{2}\left(\varepsilon_{i\alpha}-\mu_{\alpha}\right)c_{i\alpha}^{+}c_{i\alpha}\right]\right\} \]
 where $\beta=1/k_{B}T$ is the inverse temperature, electrons occupying
the two sets of Lippmann-Schwinger scattering states are characterized
by the chemical potentials $\mu_{\alpha}$ $(\alpha=1,2)$ respectively,
and $Z$ is a normalization factor ensuring that $\mbox{Tr}\left(\rho_{0}\right)=1$.
Here, $\left\{ c_{i\alpha}^{+}\right\} $ and $\left\{ c_{i\alpha}\right\} $
are the creation and annihilation operators for the complete and orthonormal
set of the Lippmann-Schwinger scattering states $\left\{ \left|\Phi_{i\alpha}\right\rangle \right\} $.
The fermion field operator $\Psi(r)$ ($\Psi^{+}(r)$) can thus be
expressed as a linear combination of $\left\{ c_{i\alpha}\right\} $
($\left\{ c_{i\alpha}^{+}\right\} $).

We now define the contour-ordered Green's function

\begin{equation}
G(rt,r^{\prime}t^{\prime})=(i\hbar)^{-1}\left\langle T_{C}\psi_{H}(rt)\psi_{H}^{+}(r^{\prime}t^{\prime})\right\rangle \label{eq:def-g}\end{equation}
where the angular bracket $\left\langle \cdots\right\rangle =\textrm{tr}\left(\rho_{0}\cdots\right)$.
By virtue of the grand-canonical structure of $\rho_{0}$, we have
thus allowed for an ensemble with a variable total number of electrons.
The contour $C$ runs from $t=-\infty$ to $t=\infty$ along the upper
branch and then returns to $t=-\infty$ along the lower branch. Here
$\psi_{H}(rt)$ and $\psi_{H}^{+}(r^{\prime}t^{\prime})$ are the
fermion field operators in the Heisenberg picture.

Parallel to the original procedure of CEID \cite{McEniry-Frederiksen-Todorov-Dundas-Horsfield-2008},
our main aim is to derive the kinetic equation for the one-electron
density matrix. Then the key quantity of interest is the lesser Green's
function $G^{<}(rt,r^{\prime}t^{\prime})=-(i\hbar)^{-1}\left\langle \psi_{H}^{+}(r^{\prime}t^{\prime})\psi_{H}(rt)\right\rangle $
since its equal-time value gives the one-electron density matrix:

\begin{equation}
\rho_{e}(r,t|r',t)=-i\hbar G^{<}(rt,r^{\prime}t)=\left\langle \psi_{H}^{+}(r^{\prime}t)\psi_{H}(rt)\right\rangle \label{eq:def-1e-DM}\end{equation}
 We first derive the equation of motion for the contour-ordered Green's
function $G(rt,r^{\prime}t^{\prime})$. Differentiating $G(rt,r^{\prime}t^{\prime})$
with respect to time arguments and then using equations of motion
for the Heisenberg operators, one obtains

\begin{equation}
\left(i\hbar\partial_{t}-h_{e}(r)\right)G(rt,r^{\prime}t^{\prime})=\delta(r-r^{\prime})\delta_{C}(t-t^{\prime})-F(r)\Gamma_{\mu}(rt,r^{\prime}t^{\prime})\label{eq:eom-g-t}\end{equation}

\begin{equation}
\left(-i\hbar\partial_{t^{\prime}}-h_{e}(r^{\prime})\right)G(rt,r^{\prime}t^{\prime})=\delta(r-r^{\prime})\delta_{C}(t-t^{\prime})-\Gamma_{\mu}^{\prime}(rt,r^{\prime}t^{\prime})F(r^{\prime})\label{eq:eom-g-tp}\end{equation}
 where $h_{e}(r)=-\frac{\hbar^{2}}{2m}\bigtriangledown^{2}+V(r)$
and $\delta_{C}(t-t^{\prime})$ is the contour delta function \cite{Rammer-Book-2007}.
Two new nonequilibrium Green's functions are introduced here \begin{equation}
\Gamma_{\mu}(rt,r^{\prime}t^{\prime})=(i\hbar)^{-1}\left\langle T_{C}X_{H}(t)\psi_{H}(rt)\psi_{H}^{+}(r^{\prime}t^{\prime})\right\rangle \label{eq:def-x-moment}\end{equation}

\begin{equation}
\Gamma_{\mu}^{\prime}(rt,r^{\prime}t^{\prime})=(i\hbar)^{-1}\left\langle T_{C}\psi_{H}(rt)\psi_{H}^{+}(r^{\prime}t^{\prime})X_{H}(t^{\prime})\right\rangle \label{eq:def-x-moment-pr}\end{equation}
 Combining equations (\ref{eq:eom-g-t}) and (\ref{eq:eom-g-tp})
gives

\begin{equation}
\dot{\rho}_{e}=\frac{1}{i\hbar}\left[h_{e},\rho_{e}\right]-\frac{1}{i\hbar}\left[F,\mu\right]\label{eq:general-ceid-1}\end{equation}
 where we used $\dot{\rho}_{e}(r,t|r',t)=-i\hbar\lim_{t\rightarrow t'}\left[\left(\partial_{t}+\partial_{t^{\prime}}\right)G(rt,r^{\prime}t^{\prime})\right]^{<}$,
and the first moment $\mu(r,t|r',t)$ is defined as:

\begin{eqnarray}
\mu(r,t|r',t) & \equiv & -i\hbar\Gamma_{\mu}^{<}(rt,r^{\prime}t)=-i\hbar\Gamma_{\mu}^{\prime<}(rt,r^{\prime}t)\nonumber \\
 & = & \left\langle \psi_{H}^{+}(r^{\prime}t)\psi_{H}(rt)X_{H}(t)\right\rangle \label{eq:def-general-mu}\end{eqnarray}

Note that the kinetic equation (\ref{eq:general-ceid-1}) always ensures
the conservation of electron number, since the relation $\dot{\left\langle N_{e}\right\rangle }=\textrm{tr}\dot{\rho}_{e}=0$
holds due to the cyclic invariance of the trace.

We proceed to find the equations of motion for $\Gamma_{\mu}(rt,r^{\prime}t^{\prime})$:

\begin{equation}
\left(i\hbar\partial_{t}-h_{e}(r)\right)\Gamma_{\mu}(rt,r^{\prime}t^{\prime})=\delta(r-r^{\prime})\delta_{C}(t-t^{\prime})\left\langle X_{H}(t)\right\rangle -F(r)\Gamma_{\mu_{2}}(rt,r^{\prime}t^{\prime})+\frac{i\hbar}{M}\Gamma_{\lambda}(rt,r^{\prime}t^{\prime})\label{eq:eom-xm-t}\end{equation}

\begin{equation}
\left(-i\hbar\partial_{t^{\prime}}-h_{e}(r^{\prime})\right)\Gamma_{\mu}(rt,r^{\prime}t^{\prime})=\delta(r-r^{\prime})\delta_{C}(t-t^{\prime})\left\langle X_{H}(t)\right\rangle -\Gamma_{\mu_{2}}^{\prime}(rt,r^{\prime}t^{\prime})F(r^{\prime})\label{eq:eom-xm-tp}\end{equation}
 where three new nonequilibrium Green's functions are defined as:
\begin{equation}
\Gamma_{\lambda}(rt,r^{\prime}t^{\prime})=(i\hbar)^{-1}\left\langle T_{C}P_{H}(t)\psi_{H}(rt)\psi_{H}^{+}(r^{\prime}t^{\prime})\right\rangle \label{eq:def-p-moment}\end{equation}

\begin{equation}
\Gamma_{\mu_{2}}(rt,r^{\prime}t^{\prime})=(i\hbar)^{-1}\left\langle T_{C}X_{H}^{2}(t)\psi_{H}(rt)\psi_{H}^{+}(r^{\prime}t^{\prime})\right\rangle \label{eq:def-x2-moment}\end{equation}

\begin{equation}
\Gamma_{\mu_{2}}^{\prime}(rt,r^{\prime}t^{\prime})=(i\hbar)^{-1}\left\langle T_{C}X_{H}(t)\psi_{H}(rt)\psi_{H}^{+}(r^{\prime}t^{\prime})X_{H}(t^{\prime})\right\rangle \label{eq:def-x2-moment-pr}\end{equation}
 In order to obtain a closed set of equations of motion, we decouple
the higher-order Green's functions $\Gamma_{\mu_{2}}(rt,r^{\prime}t^{\prime})$
and $\Gamma_{\mu_{2}}^{\prime}(rt,r^{\prime}t^{\prime})$ as follows:

\begin{equation}
\Gamma_{\mu_{2}}(rt,r^{\prime}t^{\prime})\thickapprox C_{RR}(t)G(rt,r^{\prime}t^{\prime})\label{eq:decouple-x2-moment}\end{equation}

\begin{equation}
\Gamma_{\mu_{2}}^{\prime}(rt,r^{\prime}t^{\prime})\thickapprox i\hbar D(t,t')G(rt,r^{\prime}t^{\prime})\label{eq:decouple-x2-moment-pr}\end{equation}
 where $D(t,t')=(i\hbar)^{-1}\left\langle T_{C}X_{H}(t)X_{H}(t^{\prime})\right\rangle $
is the dressed phonon Green's function and $C_{RR}(t)=i\hbar D^{<}(t,t)$.
Using these decoupling approximations and the relation $\dot{\mu}(r,t|r',t)=-i\hbar\lim_{t\rightarrow t'}\left[\left(\partial_{t}+\partial_{t^{\prime}}\right)\Gamma_{\mu}(rt,r^{\prime}t^{\prime})\right]^{<}$,
we combine equations (\ref{eq:eom-xm-t}) and (\ref{eq:eom-xm-tp})
to yield

\begin{equation}
\dot{\mu}=\frac{1}{i\hbar}\left[h_{e},\mu\right]-\frac{1}{i\hbar}C_{RR}\left[F,\rho_{e}\right]+\frac{\lambda}{M}\label{eq:general-ceid-2}\end{equation}
 where we have applied the Langreth theorem \cite{Haug-Jauho-Book-1996,Rammer-Book-2007}
to calculate $\Gamma_{\mu_{2}}^{\prime<}(rt,r^{\prime}t')=i\hbar D^{<}(t,t')G^{<}(rt,r^{\prime}t')$,
and the first moment $\lambda(r,t|r',t)$ is defined as

\begin{eqnarray}
\lambda(r,t|r',t) & \equiv & -i\hbar\Gamma_{\lambda}^{<}(rt,r^{\prime}t)\nonumber \\
 & = & \left\langle \psi_{H}^{+}(r^{\prime}t)\psi_{H}(rt)P_{H}(t)\right\rangle \label{eq:def-general-lambda}\end{eqnarray}

We continue to derive the equation of motion for $\Gamma_{\lambda}(rt,r^{\prime}t^{\prime})$:\begin{eqnarray}
\left(i\hbar\partial_{t}-h_{e}(r)\right)\Gamma_{\lambda}(rt,r^{\prime}t^{\prime}) & = & \delta(r-r^{\prime})\delta_{C}(t-t^{\prime})\left\langle P_{H}(t)\right\rangle -i\hbar K\Gamma_{\mu}(rt,r^{\prime}t^{\prime})\nonumber \\
 &  & +\Lambda(rt,r^{\prime}t^{\prime})-F(r)\Gamma_{\lambda\mu}(rt,r^{\prime}t^{\prime})\label{eq:eom-pm-t}\end{eqnarray}

\begin{equation}
\left(-i\hbar\partial_{t^{\prime}}-h_{e}(r^{\prime})\right)\Gamma_{\lambda}(rt,r^{\prime}t^{\prime})=\delta(r-r^{\prime})\delta_{C}(t-t^{\prime})\left\langle P_{H}(t)\right\rangle -\Gamma_{\lambda\mu}^{\prime}(rt,r^{\prime}t^{\prime})F(r^{\prime})\label{eq:eom-pm-tp}\end{equation}
 where three new nonequilibrium Green's functions are defined as:

\begin{equation}
\Lambda(rt,r^{\prime}t^{\prime})=\intop dr_{0}F(r_{0})\left\langle T_{C}\psi_{H}^{+}(r_{0}t)\psi_{H}(r_{0}t)\psi_{H}(rt)\psi_{H}^{+}(r^{\prime}t^{\prime})\right\rangle \label{eq:def-corre-fun}\end{equation}

\begin{equation}
\Gamma_{\lambda\mu}(rt,r^{\prime}t^{\prime})=(i\hbar)^{-1}\left\langle T_{C}P_{H}(t)X_{H}(t)\psi_{H}(rt)\psi_{H}^{+}(r^{\prime}t^{\prime})\right\rangle \label{eq:def-px-moment}\end{equation}

\begin{equation}
\Gamma_{\lambda\mu}^{\prime}(rt,r^{\prime}t^{\prime})=(i\hbar)^{-1}\left\langle T_{C}P_{H}(t)\psi_{H}(rt)\psi_{H}^{+}(r^{\prime}t^{\prime})X_{H}(t^{\prime})\right\rangle \label{eq:def-px-moment-pr}\end{equation}
 To decouple the above higher-order Green's functions, we make the
following approximations:

\[
\Lambda(rt,r^{\prime}t^{\prime})\thickapprox\intop dr_{0}F(r_{0})\left\langle \psi_{H}^{+}(r_{0}t)\psi_{H}(r_{0}t)\right\rangle \left\langle T_{C}\psi_{H}(rt)\psi_{H}^{+}(r^{\prime}t^{\prime})\right\rangle \]

\[
-\intop dr_{0}\left\langle \psi_{H}^{+}(r_{0}t)\psi_{H}(rt)\right\rangle F(r_{0})\left\langle T_{C}\psi_{H}(r_{0}t)\psi_{H}^{+}(r^{\prime}t^{\prime})\right\rangle \]

\begin{equation}
=i\hbar\intop dr_{0}F(r_{0})\rho_{e}(r_{0},t)G(rt,r^{\prime}t^{\prime})-i\hbar\intop dr_{0}\rho_{e}(r,t|r_{0},t)F(r_{0})G(r_{0}t,r^{\prime}t^{\prime})\label{eq:decouple-corre-fun}\end{equation}

\begin{eqnarray}
\Gamma_{\lambda\mu}(rt,r^{\prime}t^{\prime}) & \thickapprox & (i\hbar)^{-1}\left\langle P_{H}(t)X_{H}(t)\right\rangle \left\langle T_{C}\psi_{H}(rt)\psi_{H}^{+}(r^{\prime}t^{\prime})\right\rangle \nonumber \\
 & = & C_{PR}(t)G(rt,r^{\prime}t^{\prime})-\frac{i\hbar}{2}G(rt,r^{\prime}t^{\prime})\label{eq:decouple-px-moment}\end{eqnarray}
 \begin{equation}
\Gamma_{\lambda\mu}^{\prime}(rt,r^{\prime}t^{\prime})\thickapprox\left\langle T_{C}P_{H}(t)X_{H}(t^{\prime})\right\rangle G(rt,r^{\prime}t^{\prime})\label{eq:decouple-px-moment-pr}\end{equation}
 where $C_{PR}(t)=\frac{1}{2}\left\langle P_{H}(t)X_{H}(t)+X_{H}(t)P_{H}(t)\right\rangle $.
Using the above decoupling approximations and the relation $\dot{\lambda}(r,t|r',t)=-i\hbar\lim_{t\rightarrow t'}\left[\left(\partial_{t}+\partial_{t^{\prime}}\right)\Gamma_{\lambda}(rt,r^{\prime}t^{\prime})\right]^{<}$,
we combine equations (\ref{eq:eom-pm-t}) and (\ref{eq:eom-pm-tp})
to arrive at

\begin{equation}
\dot{\lambda}=\frac{1}{i\hbar}\left[h_{e},\lambda\right]+\textrm{tr}(\rho_{e}F)\rho_{e}+\frac{1}{2}\left\{ F,\rho_{e}\right\} -\rho_{e}F\rho_{e}-\frac{1}{i\hbar}C_{PR}\left[F,\rho_{e}\right]-K\mu\label{eq:general-ceid-3}\end{equation}
 where the Langreth theorem is applied to calculate $\Gamma_{\lambda\mu}^{\prime<}(rt,r^{\prime}t')=\left\langle X_{H}(t')P_{H}(t)\right\rangle G^{<}(rt,r^{\prime}t')$.

Kinetic equations (\ref{eq:general-ceid-1}), (\ref{eq:general-ceid-2})
and (\ref{eq:general-ceid-3}) are identical to those derived in Refs.
\cite{Horsfield-Bowler-Fisher-Todorov-Sanchez-2005,McEniry-Frederiksen-Todorov-Dundas-Horsfield-2008},
except that an extra term $\textrm{tr}(\rho_{e}F)\rho_{e}$ appears
and a $\mu F\mu$ term disappears in equation (\ref{eq:general-ceid-3})
compared with the corresponding equation in the original CEID. The
reason for the presence of the extra $\textrm{tr}(\rho_{e}F)\rho_{e}$
term is that, in the original CEID \cite{Horsfield-Bowler-Fisher-Todorov-Sanchez-2005,McEniry-Frederiksen-Todorov-Dundas-Horsfield-2008},
the expansion was with respect to $\Delta R=R-\bar{R}$, whereas we,
for convenience, here use $X=R-R_{0}$ instead. In the original CEID,
the $\mu F\mu$ term results from higher-order corrections to the
Hartree-Fock decoupling for the two-electron density matrix in equation
(\ref{eq:decouple-corre-fun}), which, however is not considered in
the present formulation. Apart from the two differences, the present
formulation of CEID is parallel to the original CEID.

In order to have a closed set of equations, one can derive the perturbation
expansion for $C_{RR}(t)$ and $C_{PR}(t)$ by using the expression
\cite{Keldysh}

\[
\left\langle A_{H}(t)B_{H}(t)\right\rangle =\left\langle T_{C}\left[e^{-\frac{i}{\hbar}\intop_{C}H_{H_{0}}^{i}(\tau)d\tau}A_{H_{0}}(t)B_{H_{0}}(t)\right]\right\rangle \]
 It is obvious that Wick's theorem can be applied directly to the
$T_{C}$-products of $\psi_{H_{0}}$'s and $\psi_{H_{0}}^{+}$'s.
For simplicity, we shall only retain the zero-order term in the expansion.
Thus we set $C_{RR}(t)=i\hbar D_{0}^{<}(t,t)=C$ ($C$ is a constant)
and $C_{PR}(t)=\frac{1}{2}\left\langle P_{H_{0}}(t)X_{H_{0}}(t)+X_{H_{0}}(t)P_{H_{0}}(t)\right\rangle =0$.
We shall hereafter consider the bare phonon Green's function $D_{0}(t,t')$
only, instead of the dressed one $D(t,t')$.

The decoupling approximations (\ref{eq:decouple-x2-moment}), (\ref{eq:decouple-x2-moment-pr}),
(\ref{eq:decouple-corre-fun}), (\ref{eq:decouple-px-moment}) and
(\ref{eq:decouple-px-moment-pr}) are the defining approximations
in the original CEID method \cite{Horsfield-Bowler-Fisher-Todorov-Sanchez-2005}.
Their key effect is to yield single-time equations of motion. These
approximations can be well understood in the framework of diagrammatic
perturbation theory as follows. Each of them represents a subset of
diagrams in the diagrammatic perturbation expansion of the corresponding
Green's function. As shown in figure 1 where we use $\Gamma_{\mu_{2}}^{\prime}(rt,r^{\prime}t^{\prime})$
as an example, the decoupling approximation (\ref{eq:decouple-x2-moment-pr})
includes the first diagram in figure 1(a) and consequently coincides
exactly with the exact perturbation expansion at the lowest-order,
while it includes only the second diagram in figure 1(a) at the second-order. 

In the NEGF-based formulation, CEID can be systematically extended
in two possible ways. Following the standard equation-of-motion technique,
one may extend the hierarchy of coupled equations of motion for Green's
functions and then truncate the hierarchy somewhere by some sort of
decoupling procedure in which a higher-order Green's functions is
expressed approximately as a product of lower-order Green's functions.
Another possibility is to improve the decoupling approximations for
Green's functions $\Gamma_{\mu_{2}}$, $\Gamma_{\mu_{2}}^{\prime}$,
$\Lambda$, $\Gamma_{\lambda\mu}$ and $\Gamma_{\lambda\mu}^{\prime}$
by adding correction terms. For instance, the last diagram in figure
1(a) (a second-order noncrossing diagram in the exact perturbation
expansion for $\Gamma_{\mu_{2}}^{\prime}$) which is absent from figure
1(b) serves as a second-order correction term to the decoupling approximation
(\ref{eq:decouple-x2-moment-pr}) illustrated in figure 1(b). We shall
focus on the way of making corrections to CEID decoupling approximations,
since it does not result in a higher hierarchy of coupled equations
(see section \ref{sub:4th-order-comparison}).

We have thus rederived the CEID equations of motion from nonequilibrium
Green's functions and generalized them to an ensemble which allows
for a variable total number of electrons. Moreover, the present formulation
allows the key approximations in CEID to be quantified in diagrammatic
terms and provides an in-principle way to improve them.

\section{Comparison with the self-consistent Born approximation\label{sec:Comparison}}

\subsection{Weak electron-phonon coupling limit\label{sub:Weak-electron-phonon-coupling}}

The CEID equations of motion to lowest-order in $F$ read:

\begin{equation}
\dot{\rho}_{e}^{(2)}=\frac{1}{i\hbar}\left[h_{e},\rho_{e}^{(2)}\right]-\frac{1}{i\hbar}\left[F,\mu^{(1)}\right]\label{eq:weak-ep-ceid-1}\end{equation}

\begin{equation}
\dot{\mu}^{(1)}=\frac{1}{i\hbar}\left[h_{e},\mu^{(1)}\right]-\frac{1}{i\hbar}C\left[F,\rho_{e}^{(0)}\right]+\frac{\lambda^{(1)}}{M}\label{eq:weak-ep-ceid-2}\end{equation}

\begin{equation}
\dot{\lambda}^{(1)}=\frac{1}{i\hbar}\left[h_{e},\lambda^{(1)}\right]+\textrm{tr}(\rho_{e}^{(0)}F)\rho_{e}^{(0)}+\frac{1}{2}\left\{ F,\rho_{e}^{(0)}\right\} -\rho_{e}^{(0)}F\rho_{e}^{(0)}-K\mu^{(1)}\label{eq:weak-ep-ceid-3}\end{equation}
where the superscript denotes the order in the coupling strength $F$.
Note that the decoupling approximations (\ref{eq:decouple-x2-moment}),
(\ref{eq:decouple-x2-moment-pr}), (\ref{eq:decouple-corre-fun}),
(\ref{eq:decouple-px-moment}) and (\ref{eq:decouple-px-moment-pr})
are exact to lowest-order in $F$. Hence the above kinetic equations
yield the \emph{exact} $\rho_{e}^{(2)}$. According to equation (\ref{eq:def-1e-DM}),
the density matrix $\rho_{e}^{(2)}$ must correspond to the sum of
all the second-order terms (the Hartree and Fock diagrams) in the
perturbation expansion of Green's function $G(1,1^{\prime})$, which
reads

\[
G^{(2)}(1,1^{\prime})=\textrm{tr}(\rho_{e}^{(0)}F)\intop d2dt_{3}D_{0}(t_{2},t_{3})G_{0}(1,2)F(r_{2})G_{0}(2,1^{\prime})\]

\begin{equation}
+i\hbar\intop d2d3G_{0}(1,2)F(r_{2})D_{0}(t_{2},t_{3})G_{0}(2,3)F(r_{3})G_{0}(3,1^{\prime})\label{eq:exact-G-2}\end{equation}
Here we use a common short notation $(k)\equiv(r_{k}t_{k})$. Starting
from this equation, one can also easily derive equations (\ref{eq:weak-ep-ceid-1}),
(\ref{eq:weak-ep-ceid-2}) and (\ref{eq:weak-ep-ceid-3}) with the
equation-of-motion technique, and furthermore identify that the second
term in the right-hand side of equation (\ref{eq:weak-ep-ceid-3})
comes from the contribution of the Hartree term.

Recall that in SCBA the sum of the second-order terms $G_{SCBA}^{(2)}$,
i.e. the first Born approximation, involves both the Hartree and Fock
diagrams and coincides exactly with equation (\ref{eq:exact-G-2}).
Hence, in the weak electron-phonon coupling limit CEID agrees exactly
with SCBA for an arbitrary nonequilibrium state of the electron-ion
system, which extends the range of validity of the conclusion in the
previous comparison \cite{McEniry-Frederiksen-Todorov-Dundas-Horsfield-2008}
for a steady state in the energy domain.

\subsection{The fourth-order in the coupling strength\label{sub:4th-order-comparison}}

We shall go beyond the weak electron-phonon coupling limit and compare
CEID and SCBA at the fourth-order in $F$. For simplicity, Hartree-like
diagrams (with a closed fermion loop) will be excluded from our analysis.
Let us consider the fourth-order SCBA Green's function $G^{(4)}(1,1^{\prime})=A_{1}(1,1^{\prime})+A_{2}(1,1^{\prime})$
which is represented by two diagrams:

\[
A_{1}(1,1^{\prime})=i\hbar\intop d2d3G_{0}(1,2)F(r_{2})D_{0}(t_{2},t_{3})G_{0}(2,3)F(r_{3})G^{(2)}(3,1^{\prime})\]

\[
A_{2}(1,1^{\prime})=i\hbar\intop d2d3G_{0}(1,2)F(r_{2})D_{0}(t_{2},t_{3})G^{(2)}(2,3)F(r_{3})G_{0}(3,1^{\prime})\]
These diagrams are shown in figure 2(a) and (b) respectively. Here
$G^{(2)}(1,1^{\prime})=i\hbar\intop d2d3G_{0}(1,2)F(r_{2})D_{0}(t_{2},t_{3})G_{0}(2,3)F(r_{3})G_{0}(3,1^{\prime})$
since the Hartree diagram is ignored. Kinetic equations for $G^{(4)}(1,1^{\prime})$
with respect to $t_{1}$ and $t_{1}^{\prime}$ yield $\dot{\rho}_{e}^{(4)}=(i\hbar)^{-1}\left[h_{e},\rho_{e}^{(4)}\right]-(i\hbar)^{-1}\left[F,\mu^{(3)}\right]$
where the corresponding $\Gamma_{\mu}^{(3)}(1,1^{\prime})=B_{1}(1,1^{\prime})+B_{2}(1,1^{\prime})$
contains two terms:

\[
B_{1}(1,1^{\prime})=-i\hbar\intop d2D_{0}(t_{1},t_{2})G_{0}(1,2)F(r_{2})G^{(2)}(2,1^{\prime})\]

\[
B_{2}(1,1^{\prime})=-i\hbar\intop d2D_{0}(t_{1},t_{2})G^{(2)}(1,2)F(r_{2})G_{0}(2,1^{\prime})\]
 Note that $B_{1}(1,1^{\prime})$ and $B_{2}(1,1^{\prime})$ result
from the temporal derivatives of $A_{1}(1,1^{\prime})$ and $A_{2}(1,1^{\prime})$
respectively. The kinetic equations for $\Gamma_{\mu}^{(3)}(1,1^{\prime})$
read\begin{equation}
i\hbar\partial_{t_{1}}\Gamma_{\mu}^{(3)}(1,1^{\prime})=h_{e}(r_{1})\Gamma_{\mu}^{(3)}(1,1^{\prime})-F(r_{1})\left[CG^{(2)}(1,1^{\prime})+S_{1}(1,1^{\prime})\right]+i\hbar\frac{\Gamma_{\lambda}^{(3)}(1,1^{\prime})}{M}\label{eq:4th-order-dmu/dt}\end{equation}

\begin{equation}
-i\hbar\partial_{t_{1}^{\prime}}\Gamma_{\mu}^{(3)}(1,1^{\prime})=\Gamma_{\mu}^{(3)}(1,1^{\prime})h_{e}(r_{1}^{\prime})-\left[i\hbar D_{0}(t_{1},t_{1}^{\prime})G^{(2)}(1,1^{\prime})+S_{2}(1,1^{\prime})\right]F(r_{1}^{\prime})\label{eq:4th-order-dmu/dt'}\end{equation}
 where $S_{1}(1,1^{\prime})=(i\hbar)^{2}\intop d2d3D_{0}(t_{1},t_{3})D_{0}(t_{1},t_{2})G_{0}(1,2)F(r_{2})G_{0}(2,3)F(r_{3})G_{0}(3,1^{\prime})$,
$S_{2}(1,1^{\prime})=(i\hbar)^{2}\intop d2d3D_{0}(t_{1},t_{2})G_{0}(1,2)F(r_{2})G_{0}(2,3)F(r_{3})D_{0}(t_{3},t_{1}^{\prime})G_{0}(3,1^{\prime})$,
and $\Gamma_{\lambda}^{(3)}(1,1^{\prime})=C_{1}(1,1^{\prime})+C_{2}(1,1^{\prime})$
where

\[
C_{1}(1,1^{\prime})=-i\hbar\intop d2G_{0}(1,2)d_{0}(t_{1},t_{2})F(r_{2})G^{(2)}(2,1^{\prime})\]

\[
C_{2}(1,1^{\prime})=-i\hbar\intop d2d_{0}(t_{1},t_{2})G^{(2)}(1,2)F(r_{2})G_{0}(2,1^{\prime})\]
 with $d_{0}(t,t')=(i\hbar)^{-1}\left\langle T_{C}P_{H_{0}}(t)X_{H_{0}}(t')\right\rangle $.
The two terms in the square bracket in equation (\ref{eq:4th-order-dmu/dt})
(equation (\ref{eq:4th-order-dmu/dt'})) correspond to all the second-order
noncrossing diagrams in $\Gamma_{\mu_{2}}(1,1^{\prime})$ ($\Gamma_{\mu_{2}}^{\prime}(1,1^{\prime})$)
(see equations (\ref{eq:def-x2-moment}) and (\ref{eq:def-x2-moment-pr})),
while $CG^{(2)}(1,1^{\prime})$ ($i\hbar D_{0}(t_{1},t_{1}^{\prime})G^{(2)}(1,1^{\prime})$)
corresponds to the single second-order diagram in the decoupling approximation
for $\Gamma_{\mu_{2}}(1,1^{\prime})$ ($\Gamma_{\mu_{2}}^{\prime}(1,1^{\prime})$)
(see equations (\ref{eq:decouple-x2-moment}) and (\ref{eq:decouple-x2-moment-pr}))
where $S_{1}(1,1^{\prime})$ ($S_{2}(1,1^{\prime})$) is not present.
This was illustrated in figure 1, where $i\hbar D_{0}(t_{1},t_{1}^{\prime})G^{(2)}(1,1^{\prime})$
and $S_{2}(1,1^{\prime})$ in equation (\ref{eq:4th-order-dmu/dt'})
are associated with the second and third diagrams in figure 1(a) respectively.
Equations (\ref{eq:4th-order-dmu/dt}) and (\ref{eq:4th-order-dmu/dt'})
lead to

\[
\dot{\mu}^{(3)}=\frac{1}{i\hbar}\left[h_{e},\mu^{(3)}\right]-\frac{1}{i\hbar}C\left[F,\rho_{e}^{(2)}\right]+\frac{\lambda^{(3)}}{M}+\Pi_{\mu}\]
 where $\Pi_{\mu}(r,t|r',t)=F(r)S_{1}^{<}(rt,r^{\prime}t)-S_{2}^{<}(rt,r^{\prime}t)F(r^{\prime})$.
It is seen that $\Pi_{\mu}$, contributed by the diagrams $S_{1}(1,1^{\prime})$
and $S_{2}(1,1^{\prime})$, serves as a correction to the CEID equation
of motion for $\mu^{(3)}$ (cf. equation (\ref{eq:general-ceid-2})
at the third-order), so as to make CEID equivalent to SCBA at the
fourth-order in the coupling strength. However this correction no
longer involves just single-time quantities. 

One can proceed to analyze the kinetic equations for $\Gamma_{\lambda}^{(3)}(1,1^{\prime})$
in a similar manner:

\begin{eqnarray}
i\hbar\partial_{t_{1}}\Gamma_{\lambda}^{(3)}(1,1^{\prime}) & = & h_{e}(r_{1})\Gamma_{\lambda}^{(3)}(1,1^{\prime})-F(r_{1})\left[\left\langle P_{H_{0}}(t_{1})X_{H_{0}}(t_{1})\right\rangle G^{(2)}(1,1^{\prime})+S_{3}(1,1^{\prime})\right]\nonumber \\
 &  & -i\hbar\intop dr_{0}\rho_{e}^{(0)}(r_{1},t_{1}|r_{0},t_{1})F(r_{0})G^{(2)}(r_{0}t_{1},r_{1}^{\prime}t_{1}^{\prime})\nonumber \\
 &  & -i\hbar\intop dr_{0}\rho_{e}^{(2)}(r_{1},t_{1}|r_{0},t_{1})F(r_{0})G_{0}(r_{0}t_{1},r_{1}^{\prime}t_{1}^{\prime})-i\hbar K\Gamma_{\mu}^{(3)}(1,1^{\prime})\label{eq:4th-order-dlambda/dt}\end{eqnarray}

\begin{equation}
-i\hbar\partial_{t_{1}^{\prime}}\Gamma_{\lambda}^{(3)}(1,1^{\prime})=\Gamma_{\lambda}^{(3)}(1,1^{\prime})h_{e}(r_{1}^{\prime})-\left[i\hbar d_{0}(t_{1},t_{1}^{\prime})G^{(2)}(1,1^{\prime})+S_{4}(1,1^{\prime})\right]F(r_{1}^{\prime})\label{eq:4th-order-dlambda/dt'}\end{equation}
 where $S_{3}(1,1^{\prime})=(i\hbar)^{2}\intop d2d3d_{0}(t_{1},t_{3})D_{0}(t_{1},t_{2})G_{0}(1,2)F(r_{2})G_{0}(2,3)F(r_{3})G_{0}(3,1^{\prime})$
and $S_{4}(1,1^{\prime})=(i\hbar)^{2}\intop d2d3d_{0}(t_{1},t_{2})G_{0}(1,2)F(r_{2})G_{0}(2,3)F(r_{3})D_{0}(t_{3},t_{1}^{\prime})G_{0}(3,1^{\prime})$.
The two terms in the square bracket in equation (\ref{eq:4th-order-dlambda/dt})
(equation (\ref{eq:4th-order-dlambda/dt'})) correspond to all the
second-order noncrossing diagrams in $\Gamma_{\lambda\mu}(1,1^{\prime})$
($\Gamma_{\lambda\mu}^{\prime}(1,1^{\prime})$) (see equations (\ref{eq:def-px-moment})
and (\ref{eq:def-px-moment-pr})), while $\left\langle P_{H_{0}}(t_{1})X_{H_{0}}(t_{1})\right\rangle G^{(2)}(1,1^{\prime})$
($i\hbar d_{0}(t_{1},t_{1}^{\prime})G^{(2)}(1,1^{\prime})$) is the
single second-order diagram in the decoupling approximation for $\Gamma_{\lambda\mu}(1,1^{\prime})$
($\Gamma_{\lambda\mu}^{\prime}(1,1^{\prime})$) (see equations (\ref{eq:decouple-px-moment})
and (\ref{eq:decouple-px-moment-pr})) where $S_{3}(1,1^{\prime})$
($S_{4}(1,1^{\prime})$) is not present.

From equations (\ref{eq:4th-order-dlambda/dt}) and (\ref{eq:4th-order-dlambda/dt'}),
one obtains

\[
\dot{\lambda}^{(3)}=\frac{1}{i\hbar}\left[h_{e},\lambda^{(3)}\right]+\frac{1}{2}\left\{ F,\rho_{e}^{(2)}\right\} -\left(\rho_{e}^{(0)}F\rho_{e}^{(2)}+\rho_{e}^{(2)}F\rho_{e}^{(0)}\right)-K\mu^{(3)}+\Pi_{\lambda}\]
 where $\Pi_{\lambda}(r,t|r',t)=F(r)S_{3}^{<}(rt,r^{\prime}t)-S_{4}^{<}(rt,r^{\prime}t)F(r^{\prime})$
which serves as a correction to the CEID equation of motion for $\lambda^{(3)}$
(cf. equation (\ref{eq:general-ceid-3}) at the third-order). Note
that a term similar to the second term in the right-hand side of equation
(\ref{eq:weak-ep-ceid-3}) does not appear because of the exclusion
of Hartree-like diagrams.

We have thus identified explicitly the corrections to the CEID equations
of motion for $\mu^{(3)}$ and $\lambda^{(3)}$ added by SCBA. These
correction terms are contributed by the diagrams which are absent
from the decoupling approximation but are present in the complete
collection of second-order noncrossing diagrams for higher-order Green's
functions $\Gamma_{\mu_{2}}(1,1^{\prime})$, $\Gamma_{\mu_{2}}^{\prime}(1,1^{\prime})$,
$\Gamma_{\lambda\mu}(1,1^{\prime})$ and $\Gamma_{\lambda\mu}^{\prime}(1,1^{\prime})$.
After this correction, CEID becomes equivalent to SCBA at the fourth-order.
In principle, one may extend this relation between CEID and SCBA to
any order in $F$. However, the amount of diagrams increases fast
with increasing order so that it would not be easy to illustrate their
relation in higher order case.

\subsection{Large ionic mass limit\label{sub:Large-ionic-mass}}

In the limit of infinite ionic mass, the electron density matrix is
determined by equation (\ref{eq:general-ceid-1}) and

\begin{equation}
\dot{\mu}=\frac{1}{i\hbar}\left[h_{e},\mu\right]-\frac{1}{i\hbar}C\left[F,\rho_{e}\right]\label{eq:infinite-M-ceid-2}\end{equation}
 Here the constant $C$ corresponds to the equal-time classical phonon
Green's function because the oscillator with infinite mass is treated
classically (see appendix). In Ref. \cite{McEniry-Frederiksen-Todorov-Dundas-Horsfield-2008},
it was shown that these coupled equations of motion are identical
to the corresponding kinetic equations for the following effective
elastic scattering problem.

Consider non-interacting electrons linearly coupled to a single infinitely
heavy classical degree of freedom $X$, with a distribution $\chi(X)=\frac{1}{2}\left[\delta(X-\sqrt{C})+\delta(X+\sqrt{C})\right]$.
The system can be described in terms of one-electron density matrix
$\rho(X,t)$ which is governed by $i\hbar\dot{\rho}(X,t)=\left[h(X),\rho(X,t)\right]$
with the one-electron Hamiltonian $h(X)=h_{0}-FX$. Define\[
\rho_{e}(t)\equiv\int\rho(X,t)\chi(X)dX\]

\[
\mu(t)\equiv\int X\rho(X,t)\chi(X)dX\]

\[
\mu_{2}(t)\equiv\int X^{2}\rho(X,t)\chi(X)dX=C\rho_{e}\]
Then $\rho_{e}(t)$ is generated \emph{exactly} by

\begin{equation}
i\hbar\dot{\rho}_{e}=\left[h_{0},\rho_{e}\right]-\left[F,\mu\right]\label{eq:elastic-scatt-eom-1}\end{equation}

\begin{equation}
i\hbar\dot{\mu}=\left[h_{0},\mu\right]-C\left[F,\rho_{e}\right]\label{eq:elastic-scatt-eom-2}\end{equation}
 which are identical to equations (\ref{eq:general-ceid-1}) and (\ref{eq:infinite-M-ceid-2})
\cite{McEniry-Frederiksen-Todorov-Dundas-Horsfield-2008}. We now
proceed to solve equations (\ref{eq:elastic-scatt-eom-1}) and (\ref{eq:elastic-scatt-eom-2})
in integral form. It is proposed that for $t>t_{0}$ this solution
can be written as \begin{equation}
\rho_{e}(t)=\rho^{<}(t,t)\,,\qquad\mu(t)=\mu^{<}(t,t)\label{eq:elastic-scatt-def-solu}\end{equation}
 where $\rho^{<}(t,t')$ and $\mu^{<}(t,t')$ are defined as

\begin{equation}
\rho^{<}(t,t')=-(i\hbar)^{2}\int G_{X}^{+}(t,t_{0})\rho_{e}(t_{0})G_{X}^{-}(t_{0},t')\chi(X)dX\label{eq:elastic-scatt-def-rho-lesser}\end{equation}

\begin{equation}
\mu^{<}(t,t')=-(i\hbar)^{2}\int XG_{X}^{+}(t,t_{0})\rho_{e}(t_{0})G_{X}^{-}(t_{0},t')\chi(X)dX\label{eq:elastic-scatt-def-mu-lesser}\end{equation}
 with $\left(i\hbar\partial_{t}-h(X)\right)G_{X}^{\pm}(t,t')=\delta(t-t')$.
Furthermore $G_{X}^{\pm}(t,t')$ can be expressed in an iterative
form

\begin{equation}
G_{X}^{\pm}(t,t')=G_{0}^{\pm}(t,t')-X\int G_{0}^{\pm}(t,t'')F(t'')G_{X}^{\pm}(t'',t')dt''\label{eq:iterative-Gx}\end{equation}
 with $\left(i\hbar\partial_{t}-h_{0}\right)G_{0}^{\pm}(t,t')=\delta(t-t')$
and $F(t)=\theta(t-t_{0})F$. Repeating the use of equation (\ref{eq:iterative-Gx})
in equations (\ref{eq:elastic-scatt-def-rho-lesser}) and (\ref{eq:elastic-scatt-def-mu-lesser}),
one finds that

\begin{equation}
\rho^{<}(t,t')=\left(1+G^{+}\cdot\Sigma_{0}^{+}\cdot\right)\rho_{0}^{<}\left(1+\cdot\Sigma_{0}^{-}\cdot G^{-}\right)+G^{+}\cdot\Sigma_{0}^{<}\cdot G^{-}\label{eq:elastic-scatt-rho-lesser}\end{equation}

\[
\mu^{<}(t,t')=-C\int\left(\rho_{0}^{<}(t,t'')F(t'')G^{-}(t'',t')+G_{0}^{+}(t,t'')F(t'')\rho^{<}(t'',t')\right)dt''\]

\begin{equation}
=-C\int\left(\rho^{<}(t,t'')F(t'')G_{0}^{-}(t'',t')+G^{+}(t,t'')F(t'')\rho_{0}^{<}(t'',t')\right)dt''\label{eq:elastic-scatt-mu-lesser}\end{equation}
with

\[
\Sigma_{0}^{\pm}(t,t')=CF(t)G_{0}^{\pm}(t,t')F(t')\]
\[
\Sigma_{0}^{<}(t,t')=CF(t)\rho_{0}^{<}(t,t')F(t')\]

\[
\rho_{0}^{<}(t,t')=-(i\hbar)^{2}G_{0}^{+}(t,t_{0})\rho_{e}(t_{0})G_{0}^{-}(t_{0},t')\]

\begin{equation}
G^{\pm}(t,t')\equiv\int G_{X}^{\pm}(t,t')\chi(X)dX=G_{0}^{\pm}(t,t')+\int G_{0}^{\pm}(t,t_{1})\Sigma_{0}^{\pm}(t_{1},t_{2})G^{\pm}(t_{2},t')dt_{1}dt_{2}\label{eq:elastic-scatt-G}\end{equation}
 Above $A=B\cdot C$ stands for $A(t,t')=\int B(t,t'')C(t'',t')dt''$.
This convention will be used where appropriate hereafter.

In the infinite mass limit, the CEID equations of motion (\ref{eq:general-ceid-1})
and (\ref{eq:infinite-M-ceid-2}) have precisely the same form as
the kinetic equations (\ref{eq:elastic-scatt-eom-1}) and (\ref{eq:elastic-scatt-eom-2})
for the elastic scattering problem. Hence, in view of the solution
to the elastic scattering problem, we may, by analogy, suggest the
following ansatz to equations (\ref{eq:general-ceid-1}) and (\ref{eq:infinite-M-ceid-2})
in the context of NEGF (cf. equations (\ref{eq:elastic-scatt-mu-lesser})
and (\ref{eq:elastic-scatt-G})):

\begin{equation}
G_{CEID}(1,1^{\prime})=G_{0}(1,1^{\prime})+C\int G_{0}(1,2)\left[F(r_{2})G_{0}(2,3)F(r_{3})\right]G(3,1^{\prime})d2d3\label{eq:infinite-M-G}\end{equation}
 and\begin{eqnarray}
\Gamma_{\mu}(1,1^{\prime}) & = & -C\intop G_{0}(1,2)F(r_{2})G(2,1^{\prime})d2\nonumber \\
 & = & -C\intop G(1,2)F(r_{2})G_{0}(2,1^{\prime})d2\label{eq:infinite-M-Gamma-mu}\end{eqnarray}

In analogy with the treatment in section (\ref{sec:Model-and-formulation}),
one can easily verify that $\rho_{e}(r,t|r',t)=-i\hbar G_{CEID}^{<}(rt,r^{\prime}t)$
and $\mu(r,t|r',t)=-i\hbar\Gamma_{\mu}^{<}(rt,r^{\prime}t)$ are solutions
to equations (\ref{eq:general-ceid-1}) and (\ref{eq:infinite-M-ceid-2}).
In the large ionic mass limit, the CEID equations of motion are thus
exactly solvable based on a correspondence between CEID and the elastic
scattering problem. Interestingly, the Dyson equation (\ref{eq:infinite-M-G})
is consistent with the Born approximation (BA) and $G_{CEID}$ contains
only one term at each order in $F$.

To compare with SCBA, we need the mixed quantum-classical perturbation
expansion for $G_{SCBA}$ which, following the discussion in the appendix,
can be obtained by replacing the quantum phonon Green's function by
the classical phonon Green's function (\ref{eq:classical-phonon-GF})
in the SCBA Dyson equation $G_{SCBA}=G_{0}+G_{0}\cdot FD_{0}GF\cdot G$
(the Hartree-like diagrams are ignored here). In the infinite ionic
mass limit, the classical phonon Green's function (\ref{eq:classical-phonon-GF})
is a constant $C$. So the SCBA Dyson equation becomes $G_{SCBA}=G_{0}+CG_{0}\cdot FGF\cdot G$.
This equation differs from the CEID solution (\ref{eq:infinite-M-G})
from the fourth-order term onwards. For instance, $G_{SCBA}^{(4)}=2G_{CEID}^{(4)}$.
In the large ionic mass limit, the difference between CEID and SCBA
is thus precisely quantified.

\section{Conclusions\label{sec:Conclusions}}

In this paper, we have considered, using the nonequilibrium Green's
function theory, a system of noninteracting electrons linearly coupled
to a quantum oscillator. A set of kinetic equations, which determine
the one-electron density matrix, are derived with the equation-of-motion
technique. Our work establishes a rigorous connection between CEID
and NEGF, and extends the scope of CEID to a general nonequilibrium
ensemble that allows for a variable total number of electrons. By
perturbation theory, the decoupling approximations used in the CEID
methodology can be quantified in diagrammatic terms.

We have compared the limiting behavior of CEID and SCBA analytically.
In the weak electron-phonon coupling limit, they agree exactly for
a general nonequilibrium state of the system. In the large ionic mass
limit, where CEID corresponds to an elastic scattering problem and
can be solved exactly, the difference between CEID and SCBA emerges
from the fourth-order term and can be quantified. In particular, we
illustrate the connection between CEID and SCBA at the fourth-order
in the coupling strength. We find that, CEID occupies a special place
between BA and SCBA, such that CEID is simpler than SCBA but is an
improvement over BA, in that CEID conserves total number of electrons.
The lowest-order SCBA corrections to the CEID equations of motion
(see section \ref{sub:4th-order-comparison}) no longer involve just
single-time quantities. This illustrates the sense in which CEID can
be thought of as the simplest particle-number conserving approximation
that, in addition, retains just single-time quantities.

The present formulation of CEID can be extended to include multiple
quantum oscillators. The purpose of the single-oscillator model calculation
is to illustrate the analytical features of CEID in a simple way so
that an analytical comparison of CEID and SCBA can be made. Like SCBA,
the present method for CEID is not applicable to problems with strong
electron-phonon correlations, which, however, have been addressed
by another CEID scheme \cite{Stella-Meister-Fisher-Horsfield-2007}
recently.

\section*{Acknowledgments}

The author is greatly indebted to Tchavdar Todorov for suggesting
this problem, and for numerous discussions and continuous ideas. The
author thanks colleagues from the EPSRC Consortium on Modelling Non-Adiabatic
Processes in Materials with Correlated Electron-Ion Dynamics for sharing
their insights into CEID. A critical reading of the manuscript by
Lev Kantorovich and Eunan McEniry is gratefully acknowledged. This
work was supported by EPSRC under Grant No. EP/C006739/01.

\section*{Appendix}

In this appendix, we shall consider the Green's function for a system
of quantum electrons coupled to a \emph{classical} oscillator and
then show how to develop a diagrammatic perturbation expansion for
it. The system Hamiltonian takes the form (\ref{eq:model-Hamiltonian})
but $X$ and $P$ are classical variables now. The mixed quantum-classical
Green's function, as usual, can be written as

\[
G(1,1^{\prime})=(i\hbar)^{-1}\left\langle T_{C}\left[\Psi_{H_{0}}(1)\Psi_{H_{0}}^{+}(1^{\prime})e^{-\frac{i}{\hbar}\intop_{C}H_{H_{0}}^{i}(\tau)d\tau}\right]\right\rangle \]
 where the angular bracket $\left\langle \cdots\right\rangle =\intop dXdP\textrm{tr}\left(\rho_{0}\cdots\right)$.
To evaluate this Green's function, we must provide a procedure for
evaluating the average of products of classical coordinates $X$'s,
while the average of products of electronic field operators can be
evaluated by Wick's theorem.

Consider a classical oscillator with position, momentum and energy

\[
X(t)=A\cos\left(\omega t-\phi\right)\]

\[
P(t)=-A\omega M\sin\left(\omega t-\phi\right)\]

\[
E=\frac{P^{2}}{2M}+\frac{1}{2}M\omega^{2}X^{2}=\frac{1}{2}M\omega^{2}A^{2}\]
 sampled from the canonical distribution

\[
\rho(X,P)=\frac{\beta\omega}{2\pi}\textrm{e}^{-\frac{1}{2}\beta M\omega^{2}A^{2}}\]
 Let $\left\langle \ldots\right\rangle $ denote averaging over $\rho$.
Changing variables from $\left(X,P\right)$ to $\left(A,\phi\right)$
with $dXdP\rightarrow M\omega AdAd\phi$,

\[
\left\langle \ldots\right\rangle =\int_{-\infty}^{\infty}dX\int_{-\infty}^{\infty}dP\ldots\rho=\frac{\beta M\omega^{2}}{2\pi}\int_{0}^{\infty}AdA\int_{0}^{2\pi}d\phi\ldots\textrm{e}^{-\frac{1}{2}\beta M\omega^{2}A^{2}}\]
 Our aim is to establish the relation

\[
L\equiv\left\langle X(t_{1})X(t_{2})\ldots X(t_{2N})\right\rangle =R\equiv\left\langle X(t_{1})X(t_{2})\right\rangle \left\langle X(t_{3})X(t_{4})\right\rangle \ldots\left\langle X(t_{2N-1})X(t_{2N})\right\rangle \]

\begin{equation}
+\textrm{all other pairings}\label{eq:classical-wick}\end{equation}
Write

\[
X(t_{i})=\frac{A}{2}\left(\textrm{e}^{i\left(\omega t_{i}-\phi\right)}+\textrm{e}^{-i\left(\omega t_{i}-\phi\right)}\right)=\frac{a_{i}+a_{i}^{*}}{2},\quad a_{i}=A\textrm{e}^{i\left(\omega t_{i}-\phi\right)}\]
 Consider $L$. Expand and integrate over $\phi$. Only terms with
$N$ $a$'s and $N$ $a^{*}$'s survive. There are

\[
N_{L}=\frac{(2N)!}{N!N!}\]
 such terms. Each is of the form

\begin{eqnarray}
 &  & \frac{1}{4^{N}}\left(\beta M\omega^{2}\int_{0}^{\infty}dAA^{2N+1}\textrm{e}^{-\frac{1}{2}\beta M\omega^{2}A^{2}}\right)\left(\textrm{e}^{i\omega\left(t_{k_{1}}+\ldots+t_{k_{N}}\right)}\times\textrm{e}^{-i\omega\left(t_{k_{N+1}}+\ldots+t_{k_{2N}}\right)}\right)\nonumber \\
 &  & =\frac{1}{2^{N}}\frac{N!}{(\beta M\omega^{2})^{N}}\left(\textrm{e}^{i\omega\left(t_{k_{1}}+\ldots+t_{k_{N}}\right)}\times\textrm{e}^{-i\omega\left(t_{k_{N+1}}+\ldots+t_{k_{2N}}\right)}\right)\label{eq:classical-wick-L}\end{eqnarray}
 with one such term occurring in $L$ for each of the $N_{L}$ possible
groupings of the $2N$ indices into two groups of $N$, $\left\{ \left(k_{1}\ldots k_{N}\right),\left(k_{N+1}\ldots k_{2N}\right)\right\} $.

Now consider $R$. Note that the classical phonon Green's function
\begin{equation}
\left\langle X(t_{i})X(t_{j})\right\rangle =\frac{1}{\beta M\omega^{2}}\cos\omega(t_{i}-t_{j})=\frac{1}{2\beta M\omega^{2}}\left(\textrm{e}^{i\omega\left(t_{i}-t_{j}\right)}+\textrm{e}^{-i\omega\left(t_{i}-t_{j}\right)}\right)\label{eq:classical-phonon-GF}\end{equation}
 In $R$ there are \[
N_{P}=(2N-1)(2N-3)\ldots1=\frac{(2N)!}{2^{N}N!}\]
 different pairings. Each pairing contributes $2^{N}$ terms, each
of the form

\begin{equation}
\frac{1}{2^{N}}\frac{1}{(\beta M\omega^{2})^{N}}\left(\textrm{e}^{i\omega\left(t_{k_{1}}+\ldots+t_{k_{N}}\right)}\times\textrm{e}^{-i\omega\left(t_{k_{N+1}}+\ldots+t_{k_{2N}}\right)}\right)\label{eq:classical-wick-R}\end{equation}
 Thus, $R$ is composed of \[
N_{R}=\frac{(2N)!}{2^{N}N!}2^{N}=\frac{(2N)!}{N!}\]
 terms, each of the form (\ref{eq:classical-wick-R}).

By symmetry, every grouping of indices $\left\{ \left(k_{1}\ldots k_{N}\right),\left(k_{N+1}\ldots k_{2N}\right)\right\} $
that occurs in $L$ occurs in $R$ and vice versa. Further, by symmetry,
if a given grouping $\left\{ \left(k_{1}\ldots k_{N}\right),\left(k_{N+1}\ldots k_{2N}\right)\right\} $
occurs $G$ times in $R$, then every other grouping must also occur
$G$ times in $R$. There are $N_{L}$ groupings. Hence

\[
G=\frac{N_{R}}{N_{L}}=N!\]
 and thus every grouping $\left\{ \left(k_{1}\ldots k_{N}\right),\left(k_{N+1}\ldots k_{2N}\right)\right\} $
occurs $N!$ times in $R$. Using this and (\ref{eq:classical-wick-R}),
we see that in $R$ each distinct grouping occurs with a prefactor

\[
\frac{1}{2^{N}}\frac{N!}{(\beta M\omega^{2})^{N}}\]
 which is the same as the prefactor with which each grouping occurs
in $L$ (see equation (\ref{eq:classical-wick-L})). Hence $L=R$.
This relation (Wick's theorem) allows us to evaluate the mixed quantum-classical
Green's function as a perturbation expansion involving only wholly
paired nuclear coordinates.\newpage{}

\section*{Figure captions}

\begin{description}
\item [{{Figure}}] 1. (a) Zero- and second-order noncrossing diagrams
in the exact $\Gamma_{\mu_{2}}^{\prime}(rt,r^{\prime}t^{\prime})$
(see equation (\ref{eq:def-x2-moment-pr})). (b) The diagrammatic
representation of the CEID decoupling approximation for $\Gamma_{\mu_{2}}^{\prime}(rt,r^{\prime}t^{\prime})$
(see equation (\ref{eq:decouple-x2-moment-pr})). The thick (thin)
straight line represents dressed (bare) electron Green's function.
The thick (thin) wavy line represents dressed (bare) phonon Green's
function. We ignore Hartree-like diagrams and corrections to a phonon
line. 
\item [{{Figure}}] 2. The diagrammatic representation of the fourth-order
SCBA Green's function $G^{(4)}(1,1^{\prime})$. 
\end{description}
\newpage{}

\begin{figure}
\begin{centering}
\includegraphics[height=5cm]{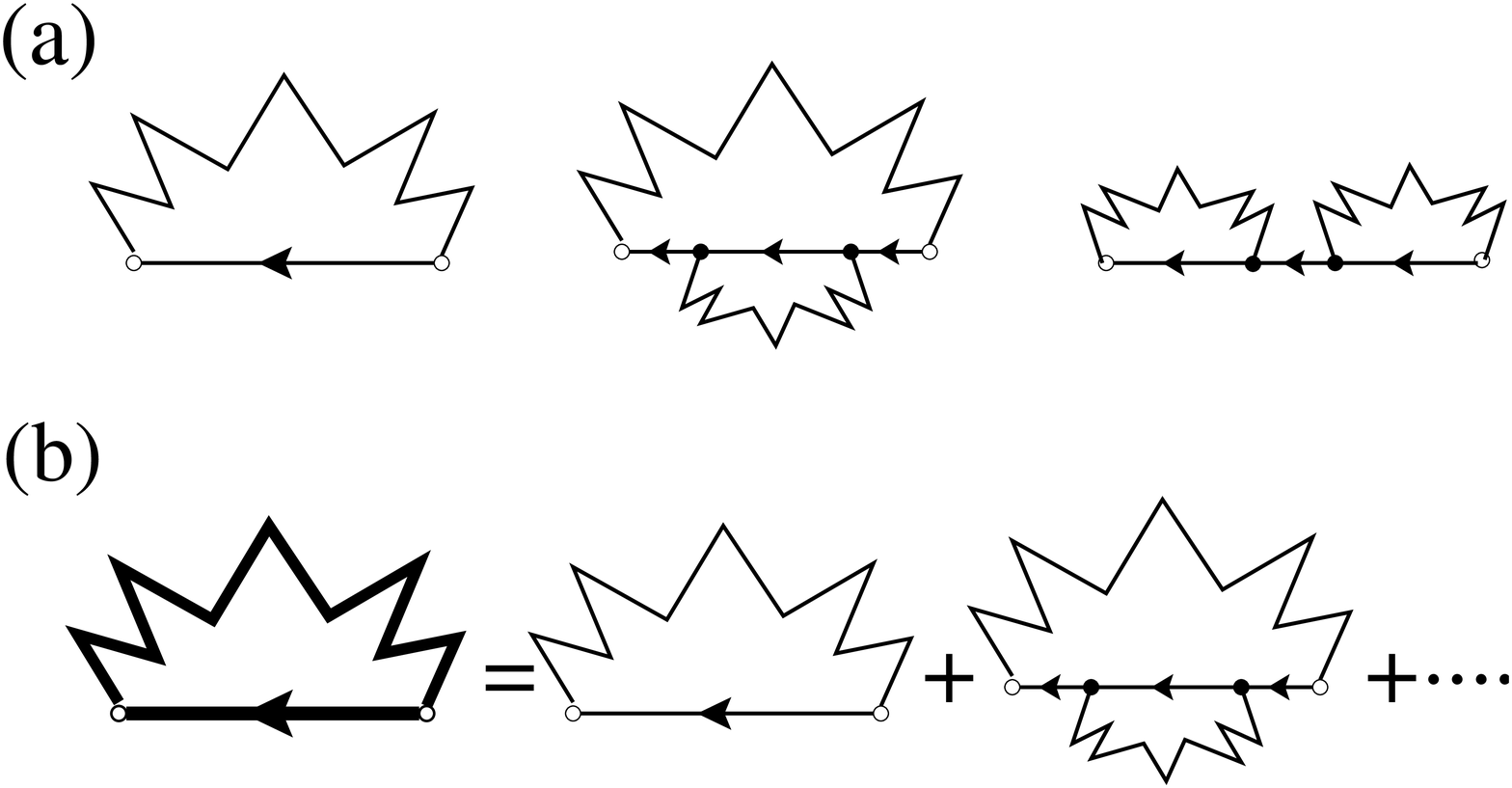} \vspace{3in}

\par\end{centering}

\centering{}Y. Wang, Figure 1 
\end{figure}

\newpage{} %
\begin{figure}
\begin{centering}
\includegraphics[height=4cm]{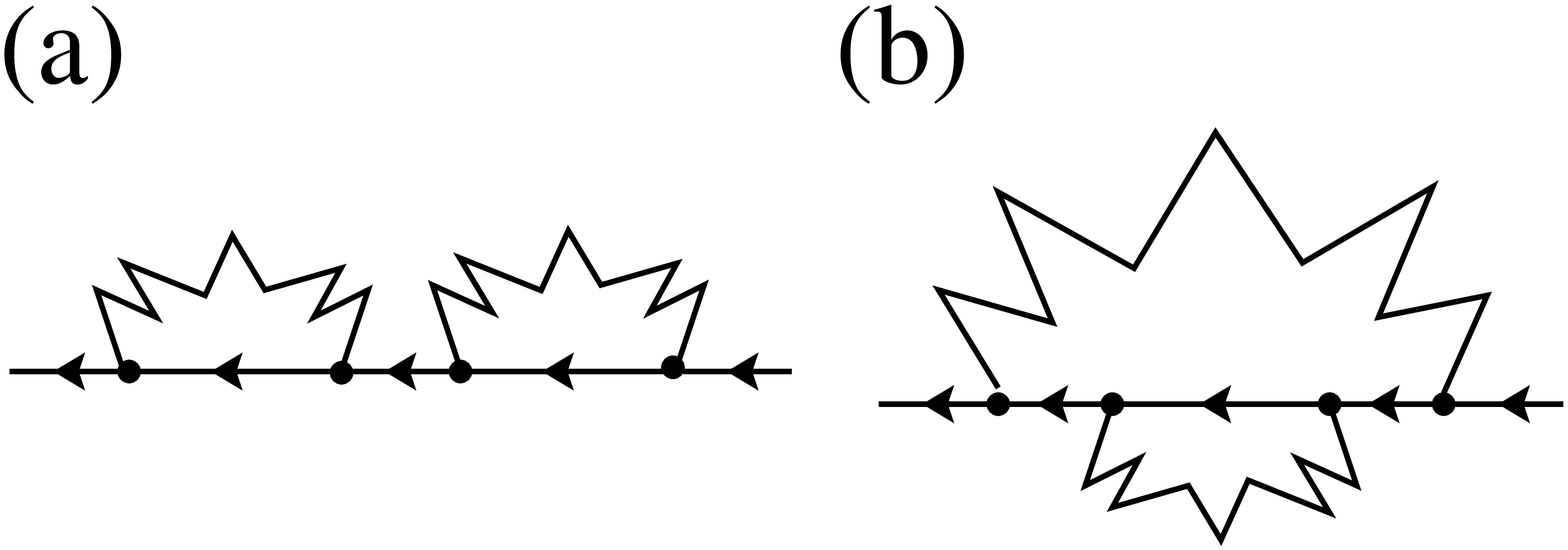} \vspace{3in}

\par\end{centering}

\centering{}Y. Wang, Figure 2 
\end{figure}

\newpage{} 
\end{document}